\documentclass{article}
\usepackage[english]{babel}
\usepackage{amsfonts,amssymb, amsmath, amsthm}
\usepackage[dvips]{graphicx}

\newcommand{\keywords}{\textbf{Key words. }\medskip}
\newcommand{\subjclass}{\textbf{MSC 2000. }\medskip}

\theoremstyle{plain}
\newtheorem{h_theorem}{Theorem}[section]

\newtheorem{h_proposition}{Proposition}[section]

\newtheorem{h_remark}{Remark}[section]

\begin{document}

\title{Separation of Variables and Integral Manifolds in One Problem
\\of Motion of Generalized Kowalevski Top}
\author{Michael P. Kharlamov and Alexander Y. Savushkin}
\date{({\it Presented by I.A.\,Lukovskii})}

\maketitle

\begin{abstract}
In the phase space of the integrable Hamiltonian system with three
degrees of freedom used to describe the motion of a Kowalevski-type
top in a double constant force field, we point out the
four-dimensional invariant manifold. It is shown that this manifold
consists of critical motions generating a smooth sheet of the
bifurcation diagram, and the induced dynamic system is Hamiltonian
with certain subset of points of degeneration of the symplectic
structure. We find the transformation separating variables in this
system. As a result, the solutions can be represented in terms of
elliptic functions of time. The corresponding phase topology is
completely described.
\end{abstract}



\section*{Introduction}
The equations of rotation of a rigid body about a fixed point in a
double constant force field have the form
$$
\begin{gathered}
\displaystyle{{\bf I} \frac{d\boldsymbol \omega} {dt}   = {\bf I}
{\boldsymbol \omega} \times {\boldsymbol \omega} + {\bf r}_1
\times {\boldsymbol \alpha } + {\bf r}_2
\times {\boldsymbol \beta },}\\[2mm]
\displaystyle{\frac{d \boldsymbol  \alpha}{dt}  = {\boldsymbol
 \alpha} \times {\boldsymbol \omega}, \quad \frac{d{\boldsymbol
\beta}}{dt}={\boldsymbol\beta}\times {\boldsymbol \omega},}
\end{gathered}
\eqno(1)
$$
where ${\bf r}_1$  and  $\ {\bf r}_2 $ are vectors immovable with
respect to the body, ${\boldsymbol \alpha }$  and  $\ {\boldsymbol
\beta }$ are vectors immovable in the inertial space, ${\bf I}$ is
the tensor of inertia at the fixed point $O$, and ${\boldsymbol
\omega }$ is the instantaneous angular velocity (all of these
objects are expressed via their components relative to certain axes
strictly attached to the body).

It is assumed that the vectors ${\bf r}_1$  and  $\ {\bf r}_2 $ have
the origin $O$. The points specified by these vectors in the moving
space are called the centers of rigging.

System (1) is a Hamiltonian system in the phase space $P^6 $
specified in ${\bf R}^9 ({\boldsymbol \omega },{\boldsymbol \alpha
},{\boldsymbol \beta })$ by geometric integrals;  $P^6 $ is
diffeomorphic to the tangent bundle $TSO(3)$.

In \cite{B1}, it is proposed to use the problem of motion of a
magnetized rigid body in gravitational and magnetic fields and the
problem of motion of a rigid body with constant distribution of
electric charge in gravitational and electric fields as physical
models of Eqs.~(1). The results obtained for system (1) in \cite{B1}
are presented in more details in  \cite{B2} within the framework of
investigation of the Euler equations on Lie algebras.

In the general case ${\bf r}_1 \times {\bf r}_2 \ne 0$  and  $
{\boldsymbol \alpha } \times {\boldsymbol \beta} \ne 0$, system (1)
without additional restrictions imposed on the parameters, unlike
the classical Euler--Poisson equations, is not reducible to a
Hamiltonian system with two degrees of freedom and does not have
known first integrals on $P^6,$ except for the energy integral
$$
H = \frac{1}{2} {\bf I} {\boldsymbol \omega} {\boldsymbol \cdot}
{\boldsymbol \omega } - {\bf r}_1 {\boldsymbol \cdot}{\boldsymbol
\alpha } - {\bf r}_2 {\boldsymbol \cdot}{\boldsymbol \beta }.
$$

In \cite{B1}, the following assumptions are introduced for system (1):
in the principal axes of the inertia tensor
$$
O{\bf e}_1 {\bf e}_2 {\bf e}_3, \eqno(2)
$$
the moments of inertia satisfy the conditions $I_1 = I_2 = 2I_3 $
and the vectors ${\bf r}_1$  and  $\ {\bf r}_2 $ are parallel to the
equatorial plane $O{\bf e}_1 {\bf e}_2 $ and mutually orthogonal.
For ${\boldsymbol \beta } = 0$, the problem reduces to the
Kowalevski integrable case of rotation of a heavy rigid body
\cite{K1}. Therefore, for the sake of brevity, the problem proposed
in \cite{B1} is called the generalized Kowalevski case and the
problem with ${\boldsymbol \beta } = 0$ is called the classical
Kowalevski case.

By the appropriate choice of measurement units and axes (2) one can
obtain
$$
{\bf I} = \operatorname{diag} \{ 2,2,1\}; \eqno(3)
$$
$$
{\bf r}_1  = {\bf e}_1,\ {\bf r}_2  = {\bf e}_2. \eqno(4)
$$

In \cite{B1}, a new general integral is indicated for the
generalized Kowalevski case. In virtue of relations (3) and (4),
this integral admits a representation:
$$
K = (\omega _1^2  - \omega _2^2  + \alpha _1  - \beta _2 )^2  +
(2\omega _1 \omega _2  + \alpha _2  + \beta _1 )^2, \eqno(5)
$$
where $\omega _i,\alpha _i,$  and  $\beta _i \;(i = 1,2,3)$ are the
components of the vectors ${\boldsymbol \omega },{\boldsymbol \alpha
}$, and ${\boldsymbol \beta }$ relative to the reference frame (2).

In \cite{Yeh}, integral (5) is generalized to the case of motion of
a gyrostat in a linear force field by supplementing a body having
property (3) with an inner rotor generating a constant moment along
the axis of dynamic symmetry. As shown, e.g, in \cite{BM}, the
component of the moment generated by potential forces introduced in
\cite{Yeh} can be reduced to the same form as in Eqs.~(1) with
property (4).

The complete Liouville integrability of the Kowalevski gyrostat in a
double force field was proved in \cite{BRS}. The Lax representation
of equations of type (1) (with gyroscopic term in the moment of
external forces, as in \cite{Yeh}) containing the spectral parameter
was obtained under conditions (3) and (4). The spectral curve of
this representation made it possible to find a new first integral
which is in involution with the corresponding generalization of
integral (5) and turns into the square of the momentum integral for
${\boldsymbol \beta } = 0$. Multi-dimensional analogs of the
Kowalevski problem were introduced in \cite{BRS}. It was proposed to
solve these problems by the method of finite-band integration. This
program was realized in \cite{BRS} for the classical Kowalevski top
and new expressions for the phase variables in the form of special
hyper-elliptic functions of time were obtained. The explicit
integration of the problem of motion of the Kowalevski top in a
double field and its qualitative and topological analysis have not
been performed yet (see also a survey in \cite{BM}).

The topological analysis of an integrable Hamiltonian system
includes the description (in certain terms) of the foliation of its
phase space into Liouville tori. In particular, this requires
finding all separating cases. These cases correspond to the points
of the bifurcation diagram of the integral map and, in the phase
space, are formed by the trajectories completely consisting of the
points at which the first integrals are not independent.

In a system with three degrees of freedom, two-dimensional Liouville
tori are, as a rule, filled with special motions corresponding to a
point of the smooth two-dimensional sheet of the bifurcation
diagram. The union of these tori over all points of the sheet is an
invariant subset of the phase space. In the neighborhood of a point
of general position, this subset is a four-dimensional manifold, and
the dynamic system induced on this subset must be Hamiltonian with
two degrees of freedom (degenerations of various kinds are expected
at the boundary of this sheet or at the points of intersection of
sheets). Thus, the invariant subsets of maximum dimension formed by
the points of dependence of integrals are specified (at least,
locally) by systems of two invariant relations of the form
$$
f_1  = 0,\ f_2  = 0. \eqno(6)
$$

The knowledge of all these systems and the analysis of the dynamics
on invariant manifolds specified by these systems is essential to
fulfil the topological analysis of the entire problem.

In the generalized Kowalevski case, we know two systems of the form
(6). The first one is obtained in \cite{B1}. Consider the manifold
    $$\{ K = 0\}  \subset P^6. \eqno(7)$$
Due to the structure of function (5), this manifold is specified by
two independent equations $Z_1 = 0$  and  $\ Z_2  = 0$. An
additional partial integral (Poisson bracket $\{ Z_1,Z_2 \}$) is
indicated. The topological analysis of the induced Hamiltonian
system with two degrees of freedom is carried out in \cite{Zot}. It
turns out that the invariant set is a four-dimensional manifold,
which is smooth everywhere but the restriction to it of the
symplectic structure degenerates on the set of zeros of the
additional integral.  This case generalizes the first Appelrot class
(Delone class) \cite{App} of motions of the classical Kowalevski
case.

The second system of the form (6) is obtained in \cite{H1}. It is
shown that, for ${\boldsymbol \beta } = 0$, the corresponding
motions transform into so-called {\it especially remarkable} motions
of the second and third Appelrot classes. The present work is
devoted to the investigation of the dynamic system on the invariant
subset indicated in \cite{H1}.

First, we make a general remark important for the sequel. The moment
of external forces ${\bf r}_1 \times {\boldsymbol \alpha } + {\bf
r}_2 \times {\boldsymbol \beta }$ appearing in (1) is preserved by
the change
$$
 \begin{pmatrix}
   \tilde {\bf r}_1  \\
   \tilde {\bf r}_2
 \end{pmatrix}  = \Theta
 \begin{pmatrix}
   {\bf r_1 }  \\
   {\bf r_2 }
   \end{pmatrix},\quad
   \begin{pmatrix}
   \tilde {\boldsymbol \alpha}  \\
   \tilde {\boldsymbol \beta}
 \end{pmatrix} = (\Theta ^T)^{-1}
 \begin{pmatrix}
   {\boldsymbol \alpha}  \\
   {\boldsymbol \beta }
 \end{pmatrix}, \eqno(8)
$$
where $\Theta $ is an arbitrary non-singular $2\times 2$ matrix.
Therefore, the {\it a-priori} assumption made in \cite{B1},
\cite{BRS} concerning the orthogonality of the radius vectors of the
centers of rigging is redundant (it suffices to require that these
centers lie in the equatorial plane of the body). This statement is
trivial enough; the possibility of orthogonalization of any pair
$({\bf r}_1,{\bf r}_2 )$ or $({\boldsymbol \alpha},{\boldsymbol
\beta })$ is indicated, e.g., in \cite{BM}. However, the authors of
\cite{BM} also indicate that, in general case, the second pair is
not orthogonal. Moreover, in \cite{B1,B2,BRS,Zot,H1}, the angle
between ${\boldsymbol \alpha }$ and ${\boldsymbol \beta }$ remains
arbitrary. This fact makes the corresponding formulas more
complicated.

Note that if the pair  $({\bf r}_1,{\bf r}_2 )$ is made orthonormal,
then there remains the arbitrary choice of $\Theta \in SO(2)$. Under
such transformation, a new pair of radius vectors of the centers of
rigging remains orthonormal and can be used as equatorial unit
vectors of the principal axes (2) to preserve properties (4). At the
same time, by choosing
$$
\Theta  =
 \begin{Vmatrix}
   {\cos \theta } & {\sin \theta }  \cr
   { - \sin \theta } & {\cos \theta }
 \end{Vmatrix},\quad
\displaystyle{ \tan 2 \theta  = \frac{2{\boldsymbol \alpha}
{\boldsymbol \cdot} {\boldsymbol \beta}} {{\boldsymbol \alpha}^2 -
{\boldsymbol \beta}^2 }},
$$
we get the orthogonal pair $(\tilde {\boldsymbol \alpha},\tilde
{\boldsymbol \beta})$.

Thus, without loss of generality, in addition to relations (4), we
can assume that the force fields are orthogonal.  This simple fact
has not been indicated yet. The elimination of the redundant
parameter makes it possible to significantly simplify all subsequent
calculations and to obtain results in a symmetric form.

\section{Invariant Subset and Its Properties}

In the sequel, we consider system (1) under assumptions (3) and (4)
with the phase space  $P^6$ specified by the formulas
$$
{\boldsymbol \alpha }^2  = a^2,\ {\boldsymbol \beta }^2  = b^2 ,\
{\boldsymbol \alpha }{\boldsymbol \cdot} {\boldsymbol \beta } =
0.\eqno(9)
$$

The case $a = b$ is singular. Indeed, as indicated in \cite{Yeh}, in
this case, there exist a group of symmetries generated by the
transformations of the configuration space and, hence, a cyclic
integral linear in the angular velocities. Therefore, for the sake
of being definite, we set
$$
a > b. \eqno(10)
$$

Denote by $G$ the general integral of the problem obtained in
\cite{BRS} and represent it in the form
$$
G = \frac{1} {4}(g_\alpha ^2  + g_\beta ^2 ) + \frac{1} {2}
\omega_3 g_\gamma - b^2 \alpha _1  - a^2 \beta _2,\eqno(11)
$$
where $g_\alpha ,\ g_\beta,$ and $g_\gamma  $ are the scalar
products of the kinetic momentum ${\bf I}{\boldsymbol \omega }$ and
the vectors ${\boldsymbol \alpha },\ {\boldsymbol \beta }$,
${\boldsymbol \alpha } \times {\boldsymbol \beta }$, respectively.

Introduce the function $F$ by setting
$$
F = (2G - p^2 H)^2  - r^4 K,
$$
where the parameters $p$ and $r$ are specified as follows:
$$
p^2  = a^2  + b^2,\ r^2  = a^2  - b^2. \eqno(12)
$$
The latter is well defined due to inequality (10). Obviously, $F$ is
a combined first integral of Eqs.~(1).

Note that the zero level of the function $F$ is specified by one of
the conditions
$$
2G - p^2 H - r^2 \sqrt K  = 0; \eqno(13)
$$
$$
2G - p^2 H + r^2 \sqrt K  = 0. \eqno(14)
$$

If ${\boldsymbol \beta } = 0$, then these conditions reduce to the
equations of the second and third Appelrot classes, respectively
\cite{App}.

Define the subset $N \subset P^6 $ as the set of critical points of
the function $F$ lying on the level $F = 0$.

The set $N$ is definitely non-empty; it contains, e.g., all points
of the form $\omega _1 = \omega _2 = 0,\;\alpha _1  - \beta _2  =
0$, and $\;\alpha _2  + \beta _1  = 0$, which are critical for $K$
and turn the expression $2G - p^2 H$ into zero.

The set $N$ is invariant under the phase flow of (1) as the set of
critical points of the general integral.

The condition $dF = 0$ means that the differentials of the functions
$H,K$, and  $G$ are linearly dependent at the points of $N$. It
immediately implies that the relation
$$
(2g - p^2 h)^2  - r^4 k = 0 \eqno(15)
$$
for the constants of these integrals is the equation of one of the
sheets of the bifurcation diagram (the investigation of this diagram
has not been completed yet) of the generalized Kowalevski top.

We use the following complex change of variables  \cite{H1} (a
generalization of the Kowalevski change; see \cite{K1}):
$$
\begin{gathered}
{x_1  = (\alpha _1  - \beta _2 ) + i(\alpha _2  + \beta _1 ),}
\quad {x_2  = \overline {x_1 },} \cr {y_1  = (\alpha _1  + \beta
_2 ) + i(\alpha _2  - \beta _1 ),} \quad {y_2  = \overline {y_1
},}\cr {z_1=\alpha _3  + i\beta _3,} \quad {z_2  = \overline {z_1
},}
  \\
w_1  = \omega _1  + i\omega _2, \quad w_2  = \overline {w_1},\quad
w_3  = \omega _3.
\end{gathered} \eqno(16)
$$

Denote the operation of differentiation with respect to the
imaginary time $it$ by primes and rewrite the equations of motion in
terms of the new variables:
$$
\begin{gathered}
{x'_1  =  - x_1 w_3  + z_1 w_1,} \quad {x'_2  = x_2 w_3  - z_2
w_2,} \cr {y'_1  =  - y_1 w_3  + z_2 w_1,} \quad {y'_2  = y_2 w_3
- z_1 w_2 ,}  \cr {2z'_1  = x_1 w_2  - y_2 w_1,} \quad {2z'_2  =
- x_2 w_1 + y_1 w_2,}
\\
2w'_1  =  - (w_1 w_3  + z_1 ),\quad 2w'_2  = w_2 w_3  + z_2, \quad
2w'_3 = y_2  - y_1.
\end{gathered} \eqno(17)
$$

Constraints (9) now take the form
$$
\begin{gathered}
z_1^2  + x_1 y_2  = r^2,\quad z_2^2  + x_2 y_1  = r^2, \\
x_1 x_2  + y_1 y_2  + 2z_1 z_2  = 2p^2.
\end{gathered}
 \eqno(18)
$$

Further on, instead of integral (11), it is convenient to consider
another general integral linearly expressed via $G$ and $H$, namely,
$$
M = \frac{1} {{r^4 }}(2G - p^2 H). \eqno(19)
$$

On the level $F = 0$, we have
$$
K = r^4 M^2. \eqno(20)
$$

In terms of variables (16), rewrite the integrals $H,\ K$, and $\ M$
as follows:
$$
\begin{gathered}
\displaystyle{H = w_1 w_2  + \frac{1} {2}w_3^2  - \frac{1} {2}(y_1
+ y_2
),\quad K = U_1 U_2,} \\
\displaystyle{M = -\frac{1} {2r^4}F_1^2+ \frac{1} {2 r^2}(U_1
+U_2),}
\end{gathered}
$$
where
$$
\begin{gathered}
\displaystyle{F_1  = \sqrt {x_1 x_2 } w_3  - \frac{1} {{\sqrt {x_1
x_2 } }}(x_2 z_1 w_1  + x_1 z_2 w_2 ),}\\
\displaystyle{U_1  = \frac{{x_2 }} {{x_1 }}(w_1^2  + x_1 ),\quad
U_2  = \frac{{x_1 }} {{x_2 }}(w_2^2  + x_2 ).}
\end{gathered}
$$

Consider the function
$$
F_2  = U_1  - U_2.
$$

\begin{h_proposition}\label{p:1}
In the domain $ x_1 x_2  \ne 0$, the invariant set $N$ is specified
by the following system of functionally independent equations:
$$
F_1  = 0,\quad F_2  = 0. \eqno(21)
$$
\end{h_proposition}

\begin{proof}
Represent relation (20) in the form
$$
\displaystyle{\bigl[F_1^2  - r^2(\sqrt {U_1 }  - \sqrt {U_2 } )^2
\bigr]\bigl[F_1^2  - r^2(\sqrt {U_1 } + \sqrt {U_2 } )^2 \bigr] =
0}, \eqno(22)
$$
where $\displaystyle{\sqrt {U_1} \  \text{and}  \ \sqrt {U_2}}$ are
chosen to be complex conjugates of each other.

On the level $F=0$, the functions $\displaystyle{F_1,\ \sqrt {U_1},
\ \text{and}\ \sqrt {U_2}}$ are independent everywhere except for
the set
$$
w_1 w_2  = 0,\;x_1  = x_2. \eqno(23)
$$

Therefore, the condition that the left-hand side of Eq.~(22) has a
critical point leads to Eqs.~(21). It is clear that points (23) also
satisfy these equations. Thus, it remains to notice that the
functions $F_1$ and $F_2$ are independent on the level $F=0$
everywhere in their domain of definition including points (23).
\end{proof}

The system of invariant relations (21) is obtained in \cite{H1}
without using the first integrals. In virtue of the above definition
and Proposition \ref {p:1}, the indicated system describes a certain
smooth four-dimensional (non-closed) manifold
$$
N^4  = N \cap \{ x_1 x_2  \ne 0\},
$$
and $N$ is the least invariant subset of $P^6 $ containing $N^4$.

\begin{h_remark}\label{r:2}\rm
It is easy to see that the invariant set $N$, as a whole, is
stratified, namely,
$$
N = \mathop  \cup \limits_{i = 1}^4 N^i,\quad\dim N^i  =
i,\quad\partial N^i  \subset \mathop  \cup \limits_{j = 1}^{i - 1}
N^j.
$$
Moreover, in virtue of Proposition \ref{p:1}, all $N^i$ with $i < 4$
belong to a subset of the phase space specified by the equation
$$
x_1 x_2  = 0 \eqno(24)
$$
(e.g.,  $N^1  = \{ w_1 w_2  = 0,w_3  = 0,x_1 x_2  = 0\}$ is
diffeomorphic to $2S^1 $). Therefore, the existence of singularity
(24) in the expressions for  $F_1$ and  $\ F_2$ is in no case
accidental. If, in relations (21), we remove the denominators, then
the set of solutions of the obtained system contains the entire
four-dimensional manifold (24). This manifold is not everywhere
critical for the function $F$ (however, $F$ is identically zero on
it). In particular, all trajectories starting from this manifold
fill a set in $P^6$, which is almost everywhere five-dimensional.
\end{h_remark}

The following statement demonstrates that if we restrict ourselves
to relations (21), i.e., study the dynamics only on $N^4$, then we
do not lose any trajectory of the dynamic system on $N$.

\begin{h_proposition}\label{p:2}
Set $(24)$ does not contain subsets invariant under the phase flow
of system~$(1)$.
\end{h_proposition}

To prove this proposition, it is necessary to compute the
derivatives of $x_1 x_2 $ in virtue of Eqs.~(17) up to the fourth
order, inclusively, and show that they cannot vanish simultaneously
on the set specified by relation (24). It is worth noting that the
indicated strong degeneration of this subset also takes place for
motions of the heavy Kowalevski top.  In that case, condition (24)
means that the axis of dynamic symmetry of the top is vertical.
Special attention is given to this phenomenon both in the classical
papers (see, e.g., \cite{App}) and in recent investigations dealing
with the computer animation of motion (see \cite{Rich} , where one
can also find an extensive bibliography of works in this field
devoted to the investigation of heavy Kowalevski tops).

\begin{h_proposition}\label{p:3}
The differential 2-form induced on the manifold $N^4$ by the
symplectic structure of the space $P^6$ providing the Hamiltonian
property of Eqs.~$(1)$ is non-degenerate everywhere except for the
subset specified by the equation $ L = 0$, where
$$
L = \frac{1} {{\sqrt {x_1 x_2 } }}\,\Bigl[w_1 w_2  + \frac{{x_1
x_2 + z_1 z_2 }} {{2r^2 }}(U_1  + U_2 )\Bigr].
$$
\end{h_proposition}

\begin{proof}
The Poisson brackets of the functions on ${\bf R}^9 ({\boldsymbol \omega
},{\boldsymbol \alpha },{\boldsymbol \beta })$ specifying the indicated
symplectic structure are determined according to the following rules \cite{B1}:
$$
\begin{gathered}
\{ g_i,g_j \}  = \varepsilon _{ijk} g_k,\quad \{ g_i,\alpha _j \}
= \varepsilon _{ijk} \alpha _k, \quad\{ g_i,\beta _j \}  =
\varepsilon _{ijk} \beta _k,\\
\{ \alpha _i,\alpha _j \}  = \{ \beta _i,\beta _j \}  = \{ \alpha
_i,\beta _j \}  = 0,
\end{gathered}
\eqno(25)
$$
where $g_1  = 2\omega _1 $, $g_2  = 2\omega _2,$ and $g_3  = \omega
_3$ are the components of the kinetic momentum.

In relations (25), we now pass to variables (16) and compute the Poisson bracket for the
functions $F_1$  and  $\ F_2 $. In view of relations (21), this gives
$$
\{ F_1,F_2 \}  =  - r^2 L.
$$

The tangent space  $T_q N^4$ is a skew-orthogonal complement of the
span of vectors included in the Hamiltonian fields with Hamiltonians
$F_1$  and  $\ F_2$. By the Cartan formula (see, e.g., \cite{Fom},
p.~231), the restriction of the symplectic structure to $T_q N^4$ is
non-degenerate provided that $\{ F_1,F_2 \}(q) \ne 0$.

\end{proof}

\begin{h_proposition}\label{p:4}
The function  $L$ is the first integral of the dynamic system
induced on $N^4$. Moreover, this integral is in involution with the
integral $M$.
\end{h_proposition}

\begin{proof} As shown in  \cite{H1}, in virtue of
(17) we can write
$$
F_1 ^\prime   = \mu _1 F_2,\quad F_2 ^\prime   = \mu _2 F_1.
$$
Here $\mu _1$  and  $\ \mu _2 $ are functions smooth in the
neighborhood of $N^4 $. In view of these equalities, apply the
Jacobi identity to the functions $H,\ F_1,$  and  $\ F_2 $ and
obtain that the double Poisson bracket $\{ H,\{ F_1,F_2 \} \} $ is a
linear combination of the functions $F_1$ and $\ F_2$. Therefore,
$L' \equiv 0$ on the set specified by relation (21).

It is shown by direct calculation that the following relation is
true under conditions (21):
$$
L^2  = 2p^2 M^2  + 2HM + 1 \eqno(26)
$$
and, therefore,
$$
L\{ L,M\}  = M\{ H,M\}  \equiv 0.
$$
It means that $\{ L,M\}  = 0$ for $L \ne 0$. Hence, by continuity,
$\{ L,M\}  = 0$ everywhere on $N^4 $.
\end{proof}

Thus, in the smooth part $N^4$ of the invariant subset $N$
completely specifying the entire dynamics on $N$, the equations of
motion of the generalized Kowalevski top define the Hamiltonian
system with two degrees of freedom with a closed subset of points of
degeneration of the symplectic structure nowhere dense in $N^4$.

\section{Analytic Solution}

By Proposition \ref{p:4}, to integrate the equations of motion in
the set $N$, we can use the integrals  $M$  and  $\ L$. The original
general integrals  $H,\ K$,  and  $\ G$ are expressed via these
integrals by using relations (19), (20), and (26).

\begin{h_theorem}\label{t:1}
On an arbitrary integral manifold
$$
J_{m,\ell }  = \{ M = m,L = \ell \}  \subset N, \eqno(27)
$$
the equations of motion are separated in the variables
$$
s_1  = \frac{{x_1 x_2  + z_1 z_2  + r^2 }} {{2\sqrt {x_1 x_2 }
}},\quad s_2  = \frac{{x_1 x_2  + z_1 z_2  - r^2 }} {{2\sqrt {x_1
x_2 } }} \eqno(28)
$$
and take the form
$$
\begin{gathered}
\displaystyle{s'_1  = \sqrt {s_1^2  - a^2 } \sqrt {ms_1^2  - \ell
s_1 +
\frac{1} {{4m}}(\ell ^2  - 1)},}  \\
\displaystyle{s'_2  = \sqrt {s_2^2  - b^2 } \sqrt {ms_2^2  - \ell
s_2  + \frac{1} {{4m}}(\ell ^2  - 1)}.}
\end{gathered}
\eqno(29)
$$
\end{h_theorem}

\begin{proof} In virtue of the first equation in (21), the function $M$ takes
the following form on $N$:
$$
M = \frac{1} {{2r^2 }}(U_1  + U_2 ).
$$

In virtue of the second equation in (21), we get $U_1 = U_2$.
Therefore, the integral equation $M=m$ yields
$$
U_1  = r^2 m \quad {\rm and} \quad U_2  = r^2 m. \eqno(30)
$$

Determine  $w_3 $ from the first equation in (21) and $w_1$  and $\
w_2 $ from Eqs.~(30). We obtain
$$
\displaystyle{w_3  = \frac{{z_1 w_1 }} {{x_1 }} + \frac{{z_2 w_2 }} {{x_2
}},\quad w_2  = \sqrt {\frac{{x_2 }} {{x_1 }}} R_1,\quad w_1 = \sqrt
{\frac{{x_1 }} {{x_2 }}} R_2,} \eqno(31)
$$
where
$$
R_1  = \sqrt {r^2 m - x_1 }\quad {\rm and} \quad R_2  = \sqrt {r^2 m - x_2 }.
\eqno(32)
$$

Substituting these quantities in the equation of the integral $L$,
we obtain
$$
m(x^2  + z^2 ) - \ell x + \sqrt {r^4 m^2  - 2r^2 mx\cos \sigma  +
x^2 }  = 0. \eqno(33)
$$

The variables  $x,z$,  and  $\sigma $ are defined as follows
$$
x^2  = x_1 x_2,\quad z^2  = z_1 z_2,\quad x_1  + x_2  = 2x\cos
\sigma, \eqno(34)
$$
and the radical in (33) corresponds to $w_1 w_2$, and therefore is
non-negative. The other radicals used above, including  $R_1$  and
$\ R_2,$ are algebraic.

Equation (33) now yields
$$
\begin{array}{c}
R_1 R_2  = \ell x - m(x^2  + z^2 ), \\ \\
\displaystyle{R_1^2  + R_2^2  = \frac{1} {{r^2 m}}\{ [\ell x - m(x^2 + z^2 )]^2
- x^2 \} + r^2 m,}
\end{array}
$$
Introducing the polynomial
$$
\Phi (s) = 4ms^2  - 4\ell s + \frac{1} {m}(\ell ^2  - 1),
$$
we can write in terms of variables (28)
$$
R_1  + R_2  = \frac{r} {{s_1  - s_2 }}\sqrt {\Phi (s_2 )}\quad {\rm and} \quad
R_1 - R_2  = \frac{r} {{s_1  - s_2 }}\sqrt {\Phi (s_1 )}.
\eqno(35)
$$

Using constraints (18) and relations (34), we obtain
$$
(z_1  \pm z_2 )^2  = \frac{1} {{r^2 }}[(x^2  + z^2  \pm r^2 )^2  -
2x^2 (p^2  \pm r^2 )].
$$
Hence, in terms of variables (28),
$$ z_1  + z_2
= \frac{{2r}} {{s_1  - s_2 }}\sqrt {s_1^2  - a^2 } ,\quad z_1  - z_2  =
\frac{{2r}} {{s_1  - s_2 }}\sqrt {s_2^2  - b^2 }. \eqno(36)
$$

We now differentiate relations (28) in virtue of system (17). In
view of (31), we get
$$
s'_1  = \frac{{r^2 }} {{4x^2 }}(z_1  + z_2 )(R_1  - R_2 )\quad {\rm and} \quad
s'_2 = \frac{{r^2 }} {{4x^2 }}(z_1  - z_2 )(R_1  + R_2 ).
$$

Substituting expressions (35) and (36) in these equalities, we
arrive at system (29).
\end{proof}

\begin{h_remark}\label{r:1}\rm
It is clear that the deduced equations can be integrated in elliptic
functions of time. By using the standard procedure, the solutions
are expressed in terms of Jacobi functions. Their specific form
depends on the location of the roots of the polynomials under the
radicals on the right-hand sides. The bifurcation solutions of
systems of this type correspond to stationary points of one of the
equations, i.e., to the cases for which the polynomial
$$
(s^2  - a^2 )(s^2  - b^2 )\Phi (s) \eqno(37)
$$
\end{h_remark}
\noindent possesses a multiple root.

For dimension reasons, the original phase variables on manifold (27)
are expressed via  $s_1$ and $s_2 $, though, in general, these
expressions might be multi-valued functions. We now show that the
latter have a fairly simple algebraic form.

Introduce the following notation:
$$
\begin{array}{ll}
\displaystyle{S_1 = \sqrt {s_1^2 - a^2},} &
\displaystyle{\varphi_1 = \sqrt {- \Phi (s_1)},} \\ \\

\displaystyle{S_2 = \sqrt {b^2 - s_2^2},} &
\displaystyle{\varphi_2 = \sqrt {\Phi (s_2)};}
\end{array}
\eqno(38)
$$
$$
\displaystyle{\psi  = 4ms_1 s_2  - 2\ell (s_1  + s_2 ) + \frac{1}
{m}(\ell ^2 - 1).} \eqno(39)
$$

\begin{h_theorem}\label{t:2}
On the common level of the first integrals $(27)$, by using notation
$(38)$, $(39)$, the phase variables of the generalized Kowalevski
case can be expressed, in terms of variables $(28)$, as follows:
\begin{gather*}
\displaystyle{\alpha _1  = \frac{\mathstrut 1} {{2(s_1  - s_2 )^2
}}[(s_1 s_2
- a^2 )\psi + S_1 S_2 \varphi _1 \varphi _2 ], }\\
\displaystyle{\alpha _2  = \frac{\mathstrut 1} {{2(s_1  - s_2 )^2
}}[(s_1 s_2
- a^2)\varphi _1 \varphi _2  - S_1 S_2 \psi ], } \\
\displaystyle{\beta _1  =  - \frac{\mathstrut 1} {{2(s_1  - s_2
)^2 }}[(s_1
s_2  - b^2)\varphi _1 \varphi _2  - S_1 S_2 \psi ], }\\
\displaystyle{\beta _2  = \frac{\mathstrut 1} {{2(s_1  - s_2 )^2
}}[(s_1 s_2 - b^2 )\psi + S_1 S_2 \varphi _1 \varphi _2 ], }
\tag{40}\\
 \displaystyle{\alpha _3  = \frac{\mathstrut r} {{s_1  -
s_2 }}S_1 ,\quad
\beta _3  =\frac{r} {{s_1  - s_2 }}S_2, }\\
\displaystyle{\omega _1  = \frac{\mathstrut r} {{2(s_1  - s_2
)}}(\ell - 2ms_1 )\varphi _2,\quad \omega _2  = \frac{r} {{2(s_1 -
s_2
)}}(\ell  - 2ms_2)\varphi _1, }\\
\displaystyle{\omega _3  = \frac{\mathstrut 1} {{s_1  - s_2 }}(S_2
\varphi _1 - S_1 \varphi _2 )}.
\end{gather*}
\end{h_theorem}

\begin{proof}
By using  notation (12), we represent the compatibility conditions
of constraints (18) in the variables $x$ and $z$ as follows:
$$
x^2  + z^2  + r^2  \geqslant 2a\left| x \right|,\quad \left| {x^2
+ z^2  - r^2 } \right| \leqslant 2b\left| x \right|,
$$
whence we get the {\it natural} ranges for variables (28):
$$
s_1^2  \geqslant a^2\quad {\rm and} \quad s_2^2  \leqslant b^2. \eqno(41)
$$

Hence, rewriting Eqs.~(29) in the real form, we conclude that, for
given $m$  and  $\ \ell $, the domain of possible motions in the
plane $(s_1,s_2 )$ is determined, along with inequalities (41), by
the inequalities
$$
\Phi (s_1 ) \leqslant 0\quad {\rm and} \quad \Phi (s_2 ) \geqslant 0. \eqno(42)
$$

Thus, in particular, all values (38) are real on the trajectories of
the analyzed system. The expressions for the complex variables
$x_1,\ x_2$,\, $z_1,\ z_2$,\, $w_1,\ w_2$, and $\ w_3$ in terms of
$s_1$ and $\ s_2$ are obtained by application, in sequence, of
relations (35) with regard for (32), then (36), and, finally, (31).
After this, the variables $y_1$ and $\ y_2 $ are determined from the
first two relations in (18). By the change of variables inverse to
(16), we arrive at the required dependencies (40).
\end{proof}

Note that the value $s_1 = \infty $  is ordinary for the first
equation in (29) (because the degree of the polynomial under the
radical on the right-hand side is even). Moreover, if this value
belongs to the domain of possible motions, then it is reached during
a finite period of time. Similarly, relations (40) also do not have
singularities in this case. This can be proved by the change of
variables $s_1 \mapsto 1/s_1$. Thus, in particular, we have deduced
analytic expressions for all cases in which the trajectories pass
the set specified by relation (24). It means that we have
constructed the complete analytic solution of the problem on the
invariant set $N$.

\section{Phase Topology}
In the regular case, the integral manifold  $J_{m,\ell}$ consists of
two-dimen\-sional Liouville tori. The cases when they topologically
rearrange generate the bifurcation diagram of the system on $N$. It
is natural to study this diagram in the plane of constants of the
used integrals, i.e., as the set of critical values of the map
$$
J = M \times L:N \to {\bf R}^2. \eqno(43)
$$

\begin{h_theorem}
The bifurcation diagram of map $(43)$ is a part of the system of
straight lines
$$
\ell  =  - 2ma \pm 1,\ \ell  = 2ma \pm 1,\ \ell  =  - 2mb \pm 1,\
\ell  = 2mb \pm 1, \eqno(44)
$$
and the coordinate axes of the plane $(\ell,m)$ lying in the
half-plane $\ell \geqslant 0$ and specified by the conditions of
existence of real solutions
$$
\begin{aligned}
&\ell \geqslant \max \, (2ma - 1, - 2mb + 1),\quad m > 0;  \\
&\ell \leqslant  - 2mb + 1,\quad m < 0;  \\
&\ell  = 1,\quad m = 0.
\end{aligned}
\eqno(45)
$$

\end{h_theorem}
\begin{proof}
According to Remark \ref{r:1}, the diagram contains the discriminant
set of polynomial (37) formed by straight lines (44) (in the part
corresponding to the existence of motions).

The points of the coordinate axes in the plane $(m,\ell)$ belonging
to $J(N)$ must be included in the diagram; indeed, it can be shown
that the values $m =0$ and $\ell = 0$ are attained, in particular,
on the subsets $N^i\; (i < 4),$ where $N$ fails to be smooth (see
Remark \ref{r:2}). The analytical foundation for this inclusion is
as follows.

In Eqs.~(29), we can pass to the limit as $m \to 0$. As a result, by
using relations (26), we conclude that $\left| \ell  \right| \to 1 $
and  $\;(\ell ^2 - 1)/2m \to h$. At the same time, the degree of the
polynomials under the radicals decreases to three; the form of
solutions changes. Moreover, it is clear that $K$ equals zero on the
set $N \cap \{ M = 0\}$, i.e., the corresponding motions also belong
to the class (7). [It is worth noting that, as shown in \cite{Zot},
the restriction of symplectic structure to manifold (7) degenerates
just at the points of corresponding trajectories.] Therefore, the
value $m=0$ should be regarded as corresponding to bifurcation.

On the other hand, the integral surface $J_{m,\ell}$ is preserved by
the inversion
$$ (\alpha _3,\beta _3,\omega
_3 ) \mapsto ( - \alpha _3, - \beta _3, - \omega _3 ). $$

In relations (40), this inversion is realized either by changing the
sign of the radicals $S_1$ and $\ S_2 $ or by the substitution
$(\ell,s_1,s_2) \mapsto (-\ell,- s_1,-s_2)$. This means that
$J_{m,\ell}$ and $J_{m,-\ell}$ are the same subset of the phase
space. Therefore, we need to restrict ourselves to the values of
$\ell$ of the same sign (to be definite, we choose non-negative
values); the axis $\ell = 0$ becomes the outer boundary of the
domain of existence of motions in the plane of constants of the
integrals. In virtue of Eq.~(33), $\ell$ can equal zero only for
negative values of $m$.

Thus, in addition to (44), the diagram should also be supplemented
with the point $\{m=0$, $\ell=1\}$ and the semi-axis
$$
\{ \ell  = 0,m < 0\}. \eqno(46)
$$

By analyzing the compatibility of conditions (41) and (42), we
determine the actual region of existence of motions in the form
(45).
\end{proof}

In Fig.~1, the domains  with numbers 1--9 defined by the diagram in
the plane $(m,\ell)$ correspond to different types of the integral
surfaces (27). The motion is impossible in the shaded region.

\begin{figure}[h]
\centering
\includegraphics[width=7cm,keepaspectratio]{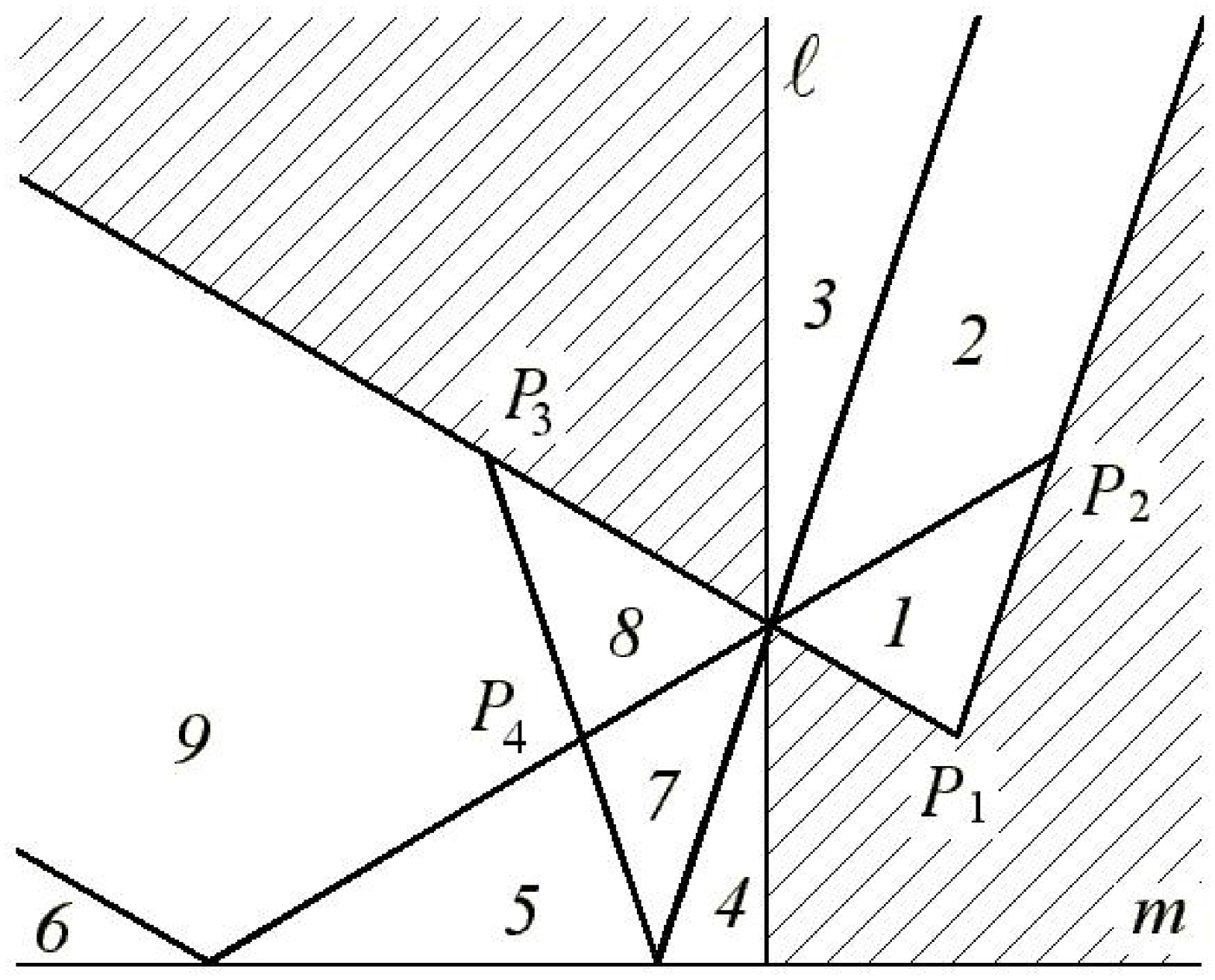}
\caption{{\small Bifurcation diagram and the domains of existence of
motions}}
\end{figure}

To determine the number of tori for the regular manifold, we note
that relations (40) give a one-valued dependence of the phase
variables on two collections of quantities
$$
(s_1,S_1,\varphi _1)\quad {\rm and} \quad (s_2,S_2,\varphi _2). \eqno(47)
$$

In this case, the signs of the radicals in (38) on each $J_{m,\ell}$
are arbitrary. However, along the trajectory, some radicals turn to
zero and then change the sign. This means that two points that only
differ by the sign of such radical lie in the same connected
component of $J_{m,\ell}$. Therefore, the number of connected
components of the regular integral manifold is equal to $2^n$, where
$n$ is the number of quantities (38) non-zero along the trajectory.
The value of $n$ is determined according to the location of roots of
polynomial (37) and does not exceed 2.

\begin{h_proposition}
Assume that the analyzed domains are numbered as indicated in
Fig.~$1$. Then the integral manifolds $J_{m,\ell}$ can be described
as follows: $\rm{a)}$ ${\bf T}^2,$ in domains $1$ and $8$, $\rm{b)}$
$2{\bf T}^2,$ in domains $2$, $7$, and $9$, and $\rm{c)}$ $4{\bf
T}^2,$ in domains $3$--$6$.
\end{h_proposition}

To determine the type of critical integral surfaces, we note that,
in each three-dimensional space of collections (47), equalities (38)
specify a pair of cylinders (elliptic or hyperbolic) with mutually
orthogonal generatrices. For the points of the straight lines (44),
a pair of cylinders corresponding to one of the variables $s_1$ or
$\ s_2$ has a tangency point. Hence, the line of their intersection
is the eight curve $S^1 \vee S^1$. Thus, on segments of the straight
lines (44) bounded by the points of lines intersection and internal
for domain (45), the integral surface consists of the components
homeomorphic to the product $S^1 \times (S^1 \vee S^1)$. Crossing
such segment, we observe one of bifurcations ${\bf T}^2  \to 2{\bf
T}^2$ typical for systems with two degrees of freedom. The number of
connected components of the form $S^1 \times (S^1 \vee S^1)$ in the
critical $J_{m,\ell}$ is determined by the number of tori in the
adjacent domains. Actually, the critical periodic trajectories (the
traces of centers of the eight curve) are motions in which one of
variables  $s_1$ or $s_2$ remains constant and equal to the multiple
root of the corresponding polynomial under the radical. In this
case, either $S_1 \equiv 0$ and $\varphi _1 \equiv 0$ or $S_2 \equiv
0$  and  $ \varphi _2 \equiv 0$. Hence, it follows from relations
(40) that  $\omega _2 = \omega _3 \equiv 0$ in the first case and
$\omega _1 = \omega _3 \equiv 0$ in the second case. The body
performs pendulum motions in which the radius vector of one of the
centers of rigging is permanently directed along the corresponding
force field.  In approaching the outer boundary of domain (45) with
the exception of the half-line (46), the tori degenerate into
circles (periodic solutions of the indicated pendulum type) and the
surfaces $S^1 \times (S^1 \vee S^1)$ degenerate into eight curves.

It is clear that the critical single-frequency motions do not appear
in the half-line (46). The corresponding bifurcation in the segments
adjacent to domains 5 and 6 is characterized by the fact that the
number of connected components of $J_{m,0}$ is half as large as at
the close regular point of the plane $(m,\ell)$. These are so-called
minimal tori. The transition from domain 4 to a segment of the
boundary set (46) is not accompanied by the decrease in the number
of components of $J_{m,\ell}$ and all cycles homotopic to a certain
marked cycle are folded so that each component covers the limiting
component twice. In a sufficiently smooth case (e.g., in the case
when $L$ is a Bott integral on the corresponding smooth level of the
integral $M$), a Klein bottle should be obtained as a result (see,
e.g., \cite{Fom}). However, according to the explicit equations
(40), this is not true in our case. Most likely, this phenomenon is
connected with the degeneration of the induced symplectic structure.

Finally, consider the nodes denoted by $P_1$--$P_4$ in Fig.~1. For
these values of the constants of integrals, each surface
$J_{m,\ell}$ contains one singular point. These points correspond to
the equilibria of the body in which both centers of rigging lie on
the corresponding axes of force fields and, hence, the moment of
forces is equal to zero. One of these points is stable: at $P_1,$
the integral surface consists of a single point. The other three
points are unstable. As indicated above, at the nodes $P_2$ and $\
P_3$ the integral surface is homeomorphic to an eight curve. At the
node $P_4$ the integral surface can be described as follows. Take a
rectangle and identify its vertices with one point; it then can be
filled with trajectories double-asymptotic to the singular point.
The boundary of this set is a bunch of four circles. This boundary
represents two pairs of pendulum motions; each pair is asymptotic to
the highest position of one of the two centers of rigging. Take four
copies of the obtained set and attach the boundary of each to the
same bunch of four circles.

All indicated phenomena are readily established by analyzing
relations (40) and the mutual location of the cylinders formed in
spaces (47).

\section*{Conclusions}

In the present work, we perform the complete investigation of
motions of the generalized Kowalevski top playing the role of
critical motions for the entire problem and generating bifurcations
of three-dimensional Liouville tori along paths crossing the sheet
specified by Eq.~(15) of the bifurcation diagram $\Sigma\subset {\bf
R}^3$ of the general integrals of the problem. Inequalities (45) are
used to deduce the equations of the boundary of a part of this sheet
corresponding to the existence of actual critical motions, i.e.,
contained in $\Sigma.$

Consider relation (22). It plays the role of the equation of the
entire integral surface in the phase space $P^6$ for the collection
of constants of integrals satisfying relation (15). It then follows,
similar to the case of the heavy Kowalevski top (the second and
third Appelrot classes), that the straight line $\{k = 0,2g=p^2h\}$
splits sheet (15) into two classes. In the first class specified by
relation (13) and corresponding to the first non-negatively definite
factor in (22), the obtained integral manifolds, being critical for
the original system, exhaust the entire corresponding integral
surface in $P^6$ as the limit of a concentric family of
three-dimensional tori and are, in this sense, stable in $P^6$. In
the second class specified by relation (14) and corresponding to the
second (hyperbolic) factor in (22), all obtained critical surfaces
in $P^6$ are hyperbolically unstable: on the same level of the first
three integrals, one can find trajectories consisting of regular
points and double-asymptotic to the corresponding two-dimensional
tori of the system on the investigated invariant set.

\end{document}